\documentclass[14pt]{revtex4}
\usepackage{amssymb}
\usepackage{latexsym}
\usepackage{epsfig}
\begin{document}                             

\title{ Emergence and expansion of cosmic space in an accelerating BIon}

\author{ Aroonkumar Beesham$^{1}$\footnote{beeshama@unizulu.ac.za}, Alireza Sepehri$^{1,2}$\footnote{alireza.sepehri3@gmail.com}}
\affiliation{$^{1}$Department of Mathematical Sciences, University of Zululand, Private Bag X1001, Kwa-Dlangezwa 3886, South Africa\\$^{2}$Research Institute for Astronomy and Astrophysics of
Maragha (RIAAM), P.O. Box 55134-441, Maragha, Iran }

\begin{abstract}
 We generalize the Padmanabhan [arXiv:hep-th/1206.4916] mechanism to an accelerating BIon and show that the difference
between the number of degrees of freedom on the boundary surface and the number of degrees
of freedom in a bulk region causes  the accelerated expansion of a BIon. We also consider the evolution of a universe which  emerges on this BIon, and obtain its Hubble parameter and energy density.
\end{abstract}

\maketitle

\noindent
{\it keywords:} Accelerating BIon; Padmanabhan mechanism; Extra dimensions; Birth of universe; Expansion of universe; Degrees of freedom

\section{Introduction}

Several years ago,  Padmanabhan suggested that the accelerated expansion of the universe is due to the difference
between the surface degrees of freedom on the holographic horizon and the bulk degrees of freedom through the simple equation
$\triangle V = \triangle t(N_{sur} - N_{bulk})$, where V is the Hubble volume, and t is the cosmic time, both expressed 
in Planck units \cite{q1}. Since then, many discussions have taken place on the Padmanabhan proposal \cite{q2,q3,q4,q5,q6,q7,q8}. For example, in one paper, with the help of this idea, the Friedmann equations of an (n + 1)-dimensional
Friedmann-Robertson-Walker universe corresponding to general relativity, Gauss-Bonnet gravity, and Lovelock gravity have been obtained  \cite{q2}. In another, the idea of treating the cosmic space as an emergent process has been applied to brane cosmology, scalar-tensor cosmology, and f(R) gravity, and the corresponding
cosmological equations in these theories have been derived \cite{q3}. In another investigation, using the Padmanabhan suggestion, the author obtained the Friedmann equations of the universe not only in four dimensional space-time and  Einstein gravity, but also in higher dimensional space-time and  other gravity theories like Gauss-Bonnet and
 Lovelock gravity with any spacial curvature \cite{q4}. 

Some other authors, have extended the evolution
equation in the Padmanabhan idea to give the Friedmann equation  in the nonflat universe corresponding to
$k = \pm 1$ by taking into account the invariant volume surrounded by the apparent horizon \cite{q5}. In another scenario, the authors showed that applying Padmanabhan's  conjecture 
to non-Einstein gravity cases encounters serious difficulties and  has
to be heavily modified to get the Friedmann equation \cite{q6}. In another paper, Ali  applied  the equations of the universe derived in the Padmanabhan model  with the corrected
entropy-area law that follows from the Generalized Uncertainty Principle (GUP) and obtained a modified
Friedmann equation due to the GUP \cite{q7}. In more recent research, the Padmanabhan idea has been constructed in a BIonic system and it was shown that all degrees of freedom inside and outside the universe are controlled by the evolution of the BIon in the extra dimensions and tend to degrees of freedom of the black F-string in string theory  \cite{q8}.

 The main question that arises is what is the origin of this inequality between the surface degrees of freedom and the bulk degrees of freedom? We answer this question in an accelerating BIon. A BIon is a configuration which has been constructed from a brane, an anti-brane and a wormhole which connects them. In our investigation, branes are expanding with   acceleration. We show that the acceleration of the branes in this BIon leads to the difference between the number of the degrees of freedom on the surface of BIon and that in a bulk. 
 
 The outline of the
paper is as  follows.  In section \ref{o1}, we will  obtain numbers of degrees of freedom on the surface and bulk. In section
\ref{o2}, we will consider the evolution of a universe which  emerges on the branes of this accelerating BIon.

\section{ Padmanabhan mechanism in an accelerating BIon}\label{o1}
In this section, we will consider the Padmanabhan mechanism in an accelerating BIon. We will show that the acceleration of the BIon leads to the difference between the number of degrees of freedom on the surface and the number in a bulk.

Previously, it has been shown that the  metric of a thermal BIon in 10-dimensional space-time is given by \cite{q9,q10}

\begin{eqnarray}
&& ds^{2} = D^{-\frac{1}{2}} H^{-\frac{1}{2}}\Big(dx_{2}^{2} + dx_{3}^{2}\Big) + D^{\frac{1}{2}} H^{-\frac{1}{2}}\Big(-f dt^{2} + dx_{1}^{2}\Big) + D^{-\frac{1}{2}} H^{\frac{1}{2}}\Big(f^{-1} dr^{2} + r^{2}d\Omega_{5}^{2}\Big)\nonumber\\&& 
\label{a1}
\end{eqnarray} 

where

\begin{eqnarray}
&& f = 1-\frac{r_{0}^{4}}{r^{4}} \quad H = 1+\frac{r_{0}^{4}\sinh^{2}\alpha}{r^{4}}\quad D = \cos^{2}\epsilon + \sin^{2}\epsilon H^{-1}
\label{a2}
\end{eqnarray}

and

\begin{eqnarray}
&& \cosh^{2} \alpha = \frac{3}{2}\frac{\cos\frac{\delta}{3} + \sqrt{3}\cos\frac{\delta}{3}}{\cos\delta}\quad \cos\epsilon = \frac{1}{\sqrt{1 + \frac{K^{2}}{r^{4}}}}
\label{a3}
\end{eqnarray}

The angle $\delta$ is defined as:

\begin{eqnarray}
&& \cos\delta = \bar{T}^{4}\sqrt{1 + \frac{K^{2}}{r^{4}}}\quad   \bar{T} = \Big(\frac{9\pi^{2}N}{4\sqrt{3}T_{D3}}\Big)^{\frac{1}{2}}T
\label{a4}
\end{eqnarray}
 
 In an accelerating BIon, the relation between the world volume of the coordinates of the accelerating D3-branes ($\tau, \sigma $) and the coordinates of 10D
Minkowski space-time ($t, r$) are \cite{q11};

\begin{eqnarray}
&& at= e^{a\sigma} \sinh(a\tau) \quad ar=e^{a\sigma} \cosh(a\tau) \quad \text{In Region I} \nonumber\\&& at= - e^{-a\sigma} \sinh(a\tau) \quad ar =  e^{-a\sigma} \cosh(a\tau) \quad \text{In Region II}
\label{a5}
\end{eqnarray}
where $a$ is the acceleration of the branes. The above equation shows that the acceleration leads to the emergence of a Rindler space-time. This space-time has two regions and the BIons in each region act in reverse to the BIons in the other region. Also, each BIon has two parts (A and B). When the branes in part A expand, the branes in part B contract, and vice versa (See figure 1).

\begin{figure*}
\includegraphics[width=10.0cm]{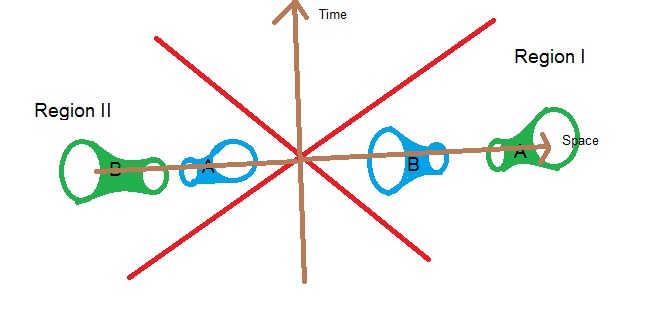}\\
\caption{ Two parts of BIons in two regions of the Rindle space-time. Parts with green color are expanding and parts with blue color are contracting. }
\end{figure*}


Substituting the results of equation (\ref{a5}) into equation (\ref{a1}), we obtain \cite{q12}:

\begin{eqnarray}
&& ds^{2}_{I,A,thermal}= D_{I-A}^{\frac{1}{2}} H_{I-A}^{-\frac{1}{2}}f_{I-A}\Big(e^{2a\sigma} + \frac{1}{\sinh^{2}(a\tau)}(\frac{dz}{d\tau})^{2} \Big)d\tau^{2} -  \nonumber\\&& D_{I-A}^{-\frac{1}{2}} H_{I-A}^{\frac{1}{2}}f_{I-A}^{-1}\Big(e^{2a\sigma}+ \frac{1}{\cosh^{2}(a\tau)}(\frac{dz}{d\sigma})^{2} \Big) d\sigma^{2} + \frac{1}{\sinh(a\tau)\cosh(a\tau)}(\frac{dz}{d\tau }\frac{dz}{d\sigma})d\tau d\sigma +  \nonumber\\&&  D_{I-A}^{-\frac{1}{2}} H_{I-A}^{\frac{1}{2}}\Big(\frac{1}{a}e^{a\sigma} \cosh(a\tau)\Big)^{2}\Big(d\theta^{2} + sin^{2}\theta d\phi^{2}\Big)  + D_{I-A}^{-\frac{1}{2}} H_{I-A}^{-\frac{1}{2}} \sum_{i=1}^{5}dx_{i}^{2}
\label{a6}
\end{eqnarray}

\begin{eqnarray}
&& ds^{2}_{II,A,thermal}= D_{II-A}^{\frac{1}{2}} H_{II-A}^{-\frac{1}{2}}f_{II-A}\Big(e^{-2a\sigma} + \frac{1}{\sinh^{2}(a\tau)}(\frac{dz}{d\tau})^{2} \Big)d\tau^{2} - \nonumber\\&& D_{II-A}^{-\frac{1}{2}} H_{II-A}^{\frac{1}{2}}f_{II-A}^{-1} \Big(e^{-2a\sigma}+ \frac{1}{\cosh^{2}(a\tau)}(\frac{dz}{d\sigma})^{2} \Big) d\sigma^{2} - \frac{1}{\sinh(a\tau)\cosh(a\tau)}(\frac{dz}{d\tau }\frac{dz}{d\sigma})d\tau d\sigma + \nonumber\\&&  D_{II-A}^{-\frac{1}{2}} H_{II-A}^{\frac{1}{2}}\Big(\frac{1}{a}e^{-a\sigma} \cosh(a\tau)\Big)^{2}\Big(d\theta^{2} + sin^{2}\theta d\phi^{2}\Big)  + D_{II-A}^{-\frac{1}{2}} H_{II-A}^{-\frac{1}{2}}\sum_{i=1}^{5}dx_{i}^{2}
\label{a7}
\end{eqnarray}

where

\begin{eqnarray}
&& f_{I-A} = 1-\frac{\Big(e^{a\sigma_{0}} \cosh(a\tau_{0})\Big)^{4}}{\Big(e^{a\sigma} \cosh(a\tau)\Big)^{4}} \quad H_{I-A} = 1+\frac{\Big(e^{a\sigma_{0}} \cosh(a\tau_{0})\Big)^{4}\sinh^{2}\alpha_{I-A}}{\Big(e^{a\sigma} \cosh(a\tau)\Big)^{4}}\nonumber\\&& D_{I-A} = \cos^{2}\epsilon_{I-A} + \sin^{2}\epsilon_{I-A} H_{I-A}^{-1}
\label{a8}
\end{eqnarray}

\begin{eqnarray}
&& f_{II-A} = 1-\frac{\Big(e^{-a\sigma_{0}} \cosh(a\tau_{0})\Big)^{4}}{\Big(e^{-a\sigma} \cosh(a\tau)\Big)^{4}} \quad H_{II-A} = 1+\frac{\Big(e^{a\sigma_{0}} \cosh(a\tau_{0})\Big)^{4}\sinh^{2}\alpha_{II-A}}{\Big(e^{a\sigma} \cosh(a\tau)\Big)^{4}}\nonumber\\&& D_{II-A} = \cos^{2}\epsilon_{II-A} + \sin^{2}\epsilon_{II-A} H_{II-A}^{-1}
\label{a9}
\end{eqnarray}

and

\begin{eqnarray}
&& \cosh^{2} \alpha_{I-A} = \frac{3}{2}\frac{\cos\frac{\delta_{I-A}}{3} + \sqrt{3}\cos\frac{\delta_{I-A}}{3}}{\cos\delta_{I-A}}\nonumber\\&& \cos\epsilon_{I-A} = \frac{1}{\sqrt{1 + \frac{K^{2}}{\Big(a^{-1}e^{-a\sigma} \cosh(a\tau)\Big)^{4}}}}
\label{a10}
\end{eqnarray}

\begin{eqnarray}
&& \cosh^{2} \alpha_{II-A} = \frac{3}{2}\frac{\cos\frac{\delta_{II-A}}{3} + \sqrt{3}\cos\frac{\delta_{II-A}}{3}}{\cos\delta_{II-A}}\nonumber\\&& \cos\epsilon_{II-A} = \frac{1}{\sqrt{1 + \frac{K^{2}}{\Big(a^{-1}e^{a\sigma} \cosh(a\tau)\Big)^{4}}}}
\label{a11}
\end{eqnarray}

The angles $\delta_{I-A}$ and $\delta_{II-A}$ are defined by:

\begin{eqnarray}
&& \cos\delta_{I-A} = \bar{T}_{0,I-A}^{4}\sqrt{1 + \frac{K^{2}}{\Big(a^{-1}e^{-a\sigma} \cosh(a\tau)\Big)^{4}}}\nonumber\\&&   \bar{T}_{0,I-A} = \Big(\frac{9\pi^{2}N}{4\sqrt{3}T_{D3}}\Big)^{\frac{1}{2}}T_{0,I-A}
\label{a12}
\end{eqnarray}

\begin{eqnarray}
&& \cos\delta_{II-A} = \bar{T}_{0,II-A}^{4}\sqrt{1 + \frac{K^{2}}{\Big(a^{-1}e^{a\sigma} \cosh(a\tau)\Big)^{4}}}\nonumber\\&&   \bar{T}_{0,II-A} = \Big(\frac{9\pi^{2}N}{4\sqrt{3}T_{D3}}\Big)^{\frac{1}{2}}T_{0,II-A}
\label{a13}
\end{eqnarray}
where $T_{0}$ is the temperature of the BIon in non-Rindler space-time.  For the above metric, the energy densities and entropies in the two regions are:

\begin{eqnarray}
&& \int d\sigma \frac{d M_{I-A}}{dz}= \int d\sigma  \frac{d M_{II-B}}{dz}=\frac{4 T_{D3}^{2}}{\pi T_{0,I-A}^{4}} \int_{\sigma_{0}}^{\infty}  d\sigma   \frac{F_{DBI,I,A}(\sigma,\tau)\Big(\frac{1}{a}e^{a\sigma} \cosh(a\tau)\Big)^{2}\Big(\sinh^{2}(a\tau)+cosh^{2}(a\tau)\Big)}{\sqrt{F_{DBI,I,A}^{2}(\sigma,\tau)-F_{DBI,I,A}^{2}(\sigma_{o},\tau)}}\times \nonumber\\&& \frac{4 \cosh^{2}\alpha_{I-A} + 1}{\cosh^{4}\alpha_{I-A}}\nonumber\\&& S_{I-A}=S_{II-B}=\frac{4 T_{D3}^{2}}{\pi T_{0,I-A}^{5}} \int_{\sigma_{0}}^{\infty}  d\sigma   \frac{F_{DBI,I,A}(\sigma,\tau)\Big(\frac{1}{a}e^{a\sigma} \cosh(a\tau)\Big)^{2}\Big(\sinh^{2}(a\tau)+cosh^{2}(a\tau)\Big)}{\sqrt{F_{DBI,I,A}^{2}(\sigma,\tau)-F_{DBI,I,A}^{2}(\sigma_{o},\tau)}}\times \nonumber\\&& \frac{4 }{\cosh^{4}\alpha_{I-A}}
\label{a14}
\end{eqnarray}

with the definition of $F_{DBI,I,A}$ given below:

\begin{eqnarray}
&& F_{DBI,I,A} = F_{DBI,II,B}= \Big(a^{-1}e^{a\sigma} \cosh(a\tau)\Big)^{2}\frac{4\cosh^{2}\alpha_{I-A} - 3}{\cosh^{4}\alpha_{I-A}}
\label{a15}
\end{eqnarray}

and

\begin{eqnarray}
&&  \int d\sigma  \frac{d M_{I-B}}{dz}= \int d\sigma  \frac{d M_{II-A}}{dz}=\frac{4 T_{D3}^{2}}{\pi T_{0,II-A}^{4}} \int_{\sigma_{0}}^{\infty}  d\sigma   \frac{F_{DBI,II,A}(\sigma,\tau)\Big(\frac{1}{a}e^{-a\sigma} \cosh(a\tau)\Big)^{2}\Big(\sinh^{2}(a\tau)+cosh^{2}(a\tau)\Big)}{\sqrt{F_{DBI,II,A}^{2}(\sigma,\tau)-F_{DBI,II,A}^{2}(\sigma_{o},\tau)}}\times \nonumber\\&& \frac{4 \cosh^{2}\alpha_{II-A} + 1}{\cosh^{4}\alpha_{II-A}}\nonumber\\&& S_{II-A}=S_{I-B}=\frac{4 T_{D3}^{2}}{\pi T_{0,II-A}^{5}} \int_{\sigma_{0}}^{\infty}  d\sigma   \frac{F_{DBI,II,A}(\sigma,\tau)\Big(\frac{1}{a}e^{-a\sigma} \cosh(a\tau)\Big)^{2}\Big(\sinh^{2}(a\tau)+cosh^{2}(a\tau)\Big)}{\sqrt{F_{DBI,II,A}^{2}(\sigma,\tau)-F_{DBI,II,A}^{2}(\sigma_{o},\tau)}}\times \nonumber\\&& \frac{4 }{\cosh^{4}\alpha_{II-A}}\label{a16}
\end{eqnarray}

with the  definition of $F_{DBI,II,A}$ given below:

\begin{eqnarray}
&& F_{DBI,II,A} = F_{DBI,I,B} =\Big(a^{-1}e^{-a\sigma} \cosh(a\tau)\Big)^{2}\frac{4\cosh^{2}\alpha_{II-A} - 3}{\cosh^{4}\alpha_{II-A}}
\label{a17}
\end{eqnarray}

Previously, it has been shown that the relation  between the entropies, energy densities and numbers of degrees  of freedom are:

\begin{eqnarray}
&& N_{sur,I-A}-N_{bulk,I-A}=\int d\sigma \frac{d M_{I-A}}{dz}= \int d\sigma  \frac{d M_{II-B}}{dz}\nonumber\\&&  N_{sur,I-A}+N_{bulk,I-A}=\frac{4\pi}{L_{P}^{2}}S_{I-A}=\frac{4\pi}{L_{P}^{2}}S_{II-B}
\label{a18}
\end{eqnarray}

\begin{eqnarray}
&& N_{sur,I-B}-N_{bulk,I-B}=\int d\sigma \frac{d M_{I-B}}{dz}= \int d\sigma  \frac{d M_{II-A}}{dz}\nonumber\\&&  N_{sur,I-B}+N_{bulk,I-B}=\frac{4\pi}{L_{P}^{2}}S_{I-B}=\frac{4\pi}{L_{P}^{2}}S_{II-A}
\label{a19}
\end{eqnarray}

Consequently, for $L_{P}^{2}=4\pi$, the numbers of degrees of freedom on the surface and in the bulk are:

\begin{eqnarray}
&& N_{sur,I-A}=\frac{4 T_{D3}^{2}}{\pi T_{0,I-A}^{4}} \int_{\sigma_{0}}^{\infty}  d\sigma   \frac{F_{DBI,I,A}(\sigma,\tau)\Big(\frac{1}{a}e^{a\sigma} \cosh(a\tau)\Big)^{2}\Big(\sinh^{2}(a\tau)+cosh^{2}(a\tau)\Big)}{\sqrt{F_{DBI,I,A}^{2}(\sigma,\tau)-F_{DBI,I,A}^{2}(\sigma_{o},\tau)}}\times \nonumber\\&& \frac{4 \cosh^{2}\alpha_{I-A} + 5}{\cosh^{4}\alpha_{I-A}}\nonumber\\&& N_{bulk,I-A}=\frac{4 T_{D3}^{2}}{\pi T_{0,I-A}^{4}} \int_{\sigma_{0}}^{\infty}  d\sigma   \frac{F_{DBI,I,A}(\sigma,\tau)\Big(\frac{1}{a}e^{a\sigma} \cosh(a\tau)\Big)^{2}\Big(\sinh^{2}(a\tau)+cosh^{2}(a\tau)\Big)}{\sqrt{F_{DBI,I,A}^{2}(\sigma,\tau)-F_{DBI,I,A}^{2}(\sigma_{o},\tau)}}\times \nonumber\\&& \frac{4 \cosh^{2}\alpha_{I-A} -3}{\cosh^{4}\alpha_{I-A}}\label{a20}
\end{eqnarray}

\begin{eqnarray}
&&  N_{sur,I-B}= \int d\sigma  \frac{d M_{II-A}}{dz}=\frac{4 T_{D3}^{2}}{\pi T_{0,II-A}^{4}} \int_{\sigma_{0}}^{\infty}  d\sigma   \frac{F_{DBI,II,A}(\sigma,\tau)\Big(\frac{1}{a}e^{-a\sigma} \cosh(a\tau)\Big)^{2}\Big(\sinh^{2}(a\tau)+cosh^{2}(a\tau)\Big)}{\sqrt{F_{DBI,II,A}^{2}(\sigma,\tau)-F_{DBI,II,A}^{2}(\sigma_{o},\tau)}}\times \nonumber\\&& \frac{4 \cosh^{2}\alpha_{II-A} + 5}{\cosh^{4}\alpha_{II-A}}\nonumber\\&& N_{bulk,I-B}= \int d\sigma  \frac{d M_{II-A}}{dz}=\frac{4 T_{D3}^{2}}{\pi T_{0,II-A}^{4}} \int_{\sigma_{0}}^{\infty}  d\sigma   \frac{F_{DBI,II,A}(\sigma,\tau)\Big(\frac{1}{a}e^{-a\sigma} \cosh(a\tau)\Big)^{2}\Big(\sinh^{2}(a\tau)+cosh^{2}(a\tau)\Big)}{\sqrt{F_{DBI,II,A}^{2}(\sigma,\tau)-F_{DBI,II,A}^{2}(\sigma_{o},\tau)}}\times \nonumber\\&& \frac{4 \cosh^{2}\alpha_{II-A} - 3}{\cosh^{4}\alpha_{II-A}}\label{a21}
\end{eqnarray}

The above equations show that the number of degrees of freedom depends on the acceleration of the branes  and the parameters of the BIon in the extra dimensions.  By increasing the acceleration, the number of degrees of freedom on the surface of one of parts increases, and number of degrees of freedom in the bulk for it correspondingly decreases, while, the number of degrees of freedom on the surface of another part decreases and number of degrees of freedom on its bulk correspondingly increases.

\section{ Evolution of the universe in an accelerating BIon}\label{o2}
In this section, we will consider the evolution of a universe which  emerges in an accelerating BIon. To this aim, we will obtain the number of degrees of freedom in terms of temperature.

Until now, various  relations for temperature of a moving system have  been proposed. For example:

\begin{eqnarray}
&& v=a\tau \nonumber\\&&\rightarrow T=\frac{T_{0}}{\sqrt{1-\frac{v^{2}}{c^{2}} }}=\frac{T_{0}}{\sqrt{1-\frac{[a\tau]^{2}}{c^{2}}} }\nonumber\\&& v=a\tau= c \sqrt{1-\frac{T_{0}^{2}}{T^{2}}}
\label{a22}
\end{eqnarray}
where $T$ is temperature of the BIon and $T_{0}$ is the critical temperature relating to the colliding point of the branes. However this relation is questionable. Based on this relation, the superconductivity phenomena depends on the system velocity!! You can move a system with special velocities to reduce its temperature to  less than that of its critical temperature, and then the system shows superconductivity by itself!! In fact, it means that a physical phenomena (superconductivity) depends on the system velocity, a result in direct conflict with the relativity law claiming that physical laws are independent of the observer velocity. This relativistic relation for temperature is not a true relation, and in fact, the temperature's relation depends on the thermocouple apparatus used. A true thermocouple rejects this definition of temperature (For example, see \cite{E1,E2,E3}). Thus, to obtain the true relation between temperature and acceleration, we make use of the concepts of the BIon:

\begin{eqnarray}
&& d M_{I-A/B} = T_{I-A/B} d S_{I-A/B} \rightarrow T_{I-A/B}=\frac{d M_{I-A/B}}{d S_{I-A/B}} 
\label{relation1}
\end{eqnarray}

Previously, thermodynamical parameters have been obtained in \cite{q12}:  

\begin{eqnarray}
&& d M_{I-A} = \frac{d M_{I-A}}{dz_{I-A}}dz_{I-A} \nonumber\\&&  d z_{I-A} = dz_{II-B}\simeq \Big(e^{-4a\sigma}\sinh^{2}(a\tau)\cosh^{2}(a\tau)\Big) \times \nonumber\\&& \Big(\frac{F_{DBI,I,A}(\tau,\sigma)\Big(\frac{F_{DBI,I,A}(\tau,\sigma)}{F_{DBI,I,A}(\tau,\sigma_{0})}-e^{-4a(\sigma-\sigma_{0})}\frac{\cosh^{2}(a\tau_{0})}{\cosh^{2}(a\tau)}\Big)^{-\frac{1}{2}}}{F_{DBI,I,A}(\tau_{0},\sigma)\Big(\frac{F_{DBI,I,A}(\tau_{0},\sigma)}{F_{DBI,I,A}(\tau_{0},\sigma_{0})}-e^{-4a(\sigma-\sigma_{0})}\frac{\cosh^{2}(a\tau_{0})}{\cosh^{2}(a\tau)}\Big)^{-\frac{1}{2}}}-\frac{\sinh^{2}(a\tau_{0})}{\sinh^{2}(a\tau)}\Big)^{-\frac{1}{2}}\nonumber\\&& \frac{d M_{I-A}}{dz}=  \frac{d M_{II-B}}{dz}=\frac{4 T_{D3}^{2}}{\pi T_{0,I-A}^{4}}  \frac{F_{DBI,I,A}(\sigma,\tau)\Big(\frac{1}{a}e^{a\sigma} \cosh(a\tau)\Big)^{2}\Big(\sinh^{2}(a\tau)+cosh^{2}(a\tau)\Big)}{\sqrt{F_{DBI,I,A}^{2}(\sigma,\tau)-F_{DBI,I,A}^{2}(\sigma_{o},\tau)}}\times \nonumber\\&& \frac{4 \cosh^{2}\alpha_{I-A} + 1}{\cosh^{4}\alpha_{I-A}}\nonumber\\&& d S_{I-A}=d S_{II-B}=\frac{4 T_{D3}^{2}}{\pi T_{0,I-A}^{5}}    \frac{F_{DBI,I,A}(\sigma,\tau)\Big(\frac{1}{a}e^{a\sigma} \cosh(a\tau)\Big)^{2}\Big(\sinh^{2}(a\tau)+cosh^{2}(a\tau)\Big)}{\sqrt{F_{DBI,I,A}^{2}(\sigma,\tau)-F_{DBI,I,A}^{2}(\sigma_{o},\tau)}}\times \nonumber\\&& \frac{4 }{\cosh^{4}\alpha_{I-A}}
\label{relation2}
\end{eqnarray}

and

\begin{eqnarray}
&&  d M_{I-B} = \frac{d M_{I-B}}{dz_{I-B}}dz_{I-B} \nonumber\\&& d z_{I-B} = d z_{II-A}\simeq \Big(e^{4a\sigma}\sinh^{2}(a\tau)\cosh^{2}(a\tau)\Big) \times \nonumber\\&& \Big(\frac{F_{DBI,II,A}(\tau,\sigma)\Big(\frac{F_{DBI,II,A}(\tau,\sigma)}{F_{DBI,II,A}(\tau,\sigma_{0})}-e^{4a(\sigma-\sigma_{0})}\frac{\cosh^{2}(a\tau_{0})}{\cosh^{2}(a\tau)}\Big)^{-\frac{1}{2}}}{F_{DBI,II,A}(\tau_{0},\sigma)\Big(\frac{F_{DBI,II,A}(\tau_{0},\sigma)}{F_{DBI,II,A}(\tau_{0},\sigma_{0})}-e^{4a(\sigma-\sigma_{0})}\frac{\cosh^{2}(a\tau_{0})}{\cosh^{2}(a\tau)}\Big)^{-\frac{1}{2}}}-\frac{\sinh^{2}(a\tau_{0})}{\sinh^{2}(a\tau)}\Big)^{-\frac{1}{2}}   \nonumber\\&&   \frac{d M_{I-B}}{dz}= \frac{d M_{II-A}}{dz}=\frac{4 T_{D3}^{2}}{\pi T_{0,II-A}^{4}}    \frac{F_{DBI,II,A}(\sigma,\tau)\Big(\frac{1}{a}e^{-a\sigma} \cosh(a\tau)\Big)^{2}\Big(\sinh^{2}(a\tau)+cosh^{2}(a\tau)\Big)}{\sqrt{F_{DBI,II,A}^{2}(\sigma,\tau)-F_{DBI,II,A}^{2}(\sigma_{o},\tau)}}\times \nonumber\\&& \frac{4 \cosh^{2}\alpha_{II-A} + 1}{\cosh^{4}\alpha_{II-A}}\nonumber\\&& d S_{II-A}= d S_{I-B}=\frac{4 T_{D3}^{2}}{\pi T_{0,II-A}^{5}}  \frac{F_{DBI,II,A}(\sigma,\tau)\Big(\frac{1}{a}e^{-a\sigma} \cosh(a\tau)\Big)^{2}\Big(\sinh^{2}(a\tau)+cosh^{2}(a\tau)\Big)}{\sqrt{F_{DBI,II,A}^{2}(\sigma,\tau)-F_{DBI,II,A}^{2}(\sigma_{o},\tau)}}\times \nonumber\\&& \frac{4 }{\cosh^{4}\alpha_{II-A}}\label{relation3}
\end{eqnarray}

Using relation (\ref{relation2} and \ref{relation3}) in relation (\ref{relation1}), we can obtain the explicit form of the temperature in an accelerating BIon as:

\begin{eqnarray}
&&  T_{I-A}= T_{0,I-A} \Big(4 \cosh^{2}\alpha_{I-A} + 1\Big)\times \nonumber\\&&\Big(e^{-4a\sigma}\sinh^{2}(a\tau)\cosh^{2}(a\tau)\Big) \times \nonumber\\&& \Big(\frac{F_{DBI,I,A}(\tau,\sigma)\Big(\frac{F_{DBI,I,A}(\tau,\sigma)}{F_{DBI,I,A}(\tau,\sigma_{0})}-e^{-4a(\sigma-\sigma_{0})}\frac{\cosh^{2}(a\tau_{0})}{\cosh^{2}(a\tau)}\Big)^{-\frac{1}{2}}}{F_{DBI,I,A}(\tau_{0},\sigma)\Big(\frac{F_{DBI,I,A}(\tau_{0},\sigma)}{F_{DBI,I,A}(\tau_{0},\sigma_{0})}-e^{-4a(\sigma-\sigma_{0})}\frac{\cosh^{2}(a\tau_{0})}{\cosh^{2}(a\tau)}\Big)^{-\frac{1}{2}}}-\frac{\sinh^{2}(a\tau_{0})}{\sinh^{2}(a\tau)}\Big)^{\frac{1}{2}}\label{relation5}
\end{eqnarray}

and

\begin{eqnarray}
&& T_{I-B}= T_{0,I-B} \Big(4 \cosh^{2}\alpha_{I-B} + 1\Big)\times \nonumber\\&&\Big(e^{4a\sigma}\sinh^{2}(a\tau)\cosh^{2}(a\tau)\Big) \times \nonumber\\&& \Big(\frac{F_{DBI,I,B}(\tau,\sigma)\Big(\frac{F_{DBI,I,B}(\tau,\sigma)}{F_{DBI,I,B}(\tau,\sigma_{0})}-e^{4a(\sigma-\sigma_{0})}\frac{\cosh^{2}(a\tau_{0})}{\cosh^{2}(a\tau)}\Big)^{-\frac{1}{2}}}{F_{DBI,I,B}(\tau_{0},\sigma)\Big(\frac{F_{DBI,I,B}(\tau_{0},\sigma)}{F_{DBI,I,B}(\tau_{0},\sigma_{0})}-e^{4a(\sigma-\sigma_{0})}\frac{\cosh^{2}(a\tau_{0})}{\cosh^{2}(a\tau)}\Big)^{-\frac{1}{2}}}-\frac{\sinh^{2}(a\tau_{0})}{\sinh^{2}(a\tau)}\Big)^{\frac{1}{2}}
\label{relation6}
\end{eqnarray}

On the other hand, previously, it has been shown that the Hubble parameter of the universe has the following relation with the number of degrees of freedom on the surface of the BIon \cite{q8}:

\begin{eqnarray}
&&  N_{sur} = \frac{4 \pi r_{A}^{2}}{L_{P}^{2}}\nonumber\\&& r_{A}=\sqrt{H^{2} + \frac{k^{2}}{\bar{a}^{2}}}
\label{a23}
\end{eqnarray}
where $H = \frac{\bar{a}}{\dot{\bar{a}}}$ is the Hubble parameter, $\bar{a}$ is the scale factor and $r_{A}$ is the apparent horizon radius for the FRW universe. Using equations (\ref{a20}, \ref{a21}, \ref{a22}, \ref{relation1}, \ref{relation2}, \ref{relation3}, \ref{relation5}, \ref{relation6} and \ref{a23}), we can obtain the Hubble parameter and scale factor of the flat universe ($k=0$) on each part of the BIon:

\begin{eqnarray}
&&  H_{I-A} \simeq \frac{313\pi^{14}k^{14}c^{14}L_{P}^{2}}{T_{D3}^{8}\sigma_{0}^{73}} \times \nonumber\\&&e^{[1-[1-\frac{\tau^{2}}{\tau_{0}^{2}}]^{1/2}\frac{T_{0}^{2}}{T^{2}}]^{8}[1-[1-\frac{\sigma^{2}}{\sigma_{0}^{2}}]^{1/2}\frac{T_{0}^{2}}{T^{2}}]^{5}}\times\nonumber\\&& [1-[1-\frac{\tau^{2}}{\tau_{0}^{2}}]^{1/2}\frac{T_{0}^{2}}{T^{2}}]^{73}\times \nonumber\\&&[1+ \frac{8\pi^{5}k^{5}c^{5}L_{P}^{2}}{T_{D3}^{313}\sigma_{0}^{313}}[1-[1-\frac{\sigma^{2}}{\sigma_{0}^{2}}]^{1/2}\frac{T_{0}^{2}}{T^{2}}]^{313}]\nonumber\\&& \bar{a}_{I-A}=e^{\int dt H_{I-A} }
\label{a24}
\end{eqnarray}

\begin{eqnarray}
&&  H_{I-B} \simeq \frac{313\pi^{14}k^{14}c^{14}L_{P}^{2}}{T_{D3}^{8}\sigma_{0}^{73}} \times \nonumber\\&&e^{-[1-[1-\frac{\tau^{2}}{\tau_{0}^{2}}]^{1/2}\frac{T_{0}^{2}}{T^{2}}]^{8}[1-[1-\frac{\sigma^{2}}{\sigma_{0}^{2}}]^{1/2}\frac{T_{0}^{2}}{T^{2}}]^{5}}\times\nonumber\\&& \frac{1}{[1-[1-\frac{\tau^{2}}{\tau_{0}^{2}}]^{1/2}\frac{T_{0}^{2}}{T^{2}}]^{73}}\times \nonumber\\&&\frac{1}{[1+ \frac{8\pi^{5}k^{5}c^{5}L_{P}^{2}}{T_{D3}^{313}\sigma_{0}^{313}}[1-[1-\frac{\sigma^{2}}{\sigma_{0}^{2}}]^{1/2}\frac{T_{0}^{2}}{T^{2}}]^{313}]}\nonumber\\&& \bar{a}_{I-A}=e^{\int dt H_{I-A} }
\label{a25}
\end{eqnarray}
The above equations show that by increasing the  temperature of the  BIon and the  acceleration of system, the  Hubble parameter and scale factor of the universe which lives on  part A of the BIon in region I decrease, while the Hubble parameter and scale factor of the universe on  part B of the BIon in region I increase. This means that by expanding one universe, another universe contracts. 
 
On the other hand, using the Friedmann equation of the flat FRW universe, we can calculate the energy density of the universe:

\begin{eqnarray}
&& \rho_{I-A}=\frac{3}{8\pi L_{P}^{2}} H_{I-A}^{2} = \nonumber\\&& \frac{3}{8\pi L_{P}^{2}}[\frac{313\pi^{14}k^{14}c^{14}L_{P}^{2}}{T_{D3}^{8}\sigma_{0}^{73}} \times \nonumber\\&&e^{[1-[1-\frac{\tau^{2}}{\tau_{0}^{2}}]^{1/2}\frac{T_{0}^{2}}{T^{2}}]^{8}[1-[1-\frac{\sigma^{2}}{\sigma_{0}^{2}}]^{1/2}\frac{T_{0}^{2}}{T^{2}}]^{5}}\times\nonumber\\&& [1-[1-\frac{\tau^{2}}{\tau_{0}^{2}}]^{1/2}\frac{T_{0}^{2}}{T^{2}}]^{73}\times \nonumber\\&&[1+ \frac{8\pi^{5}k^{5}c^{5}L_{P}^{2}}{T_{D3}^{313}\sigma_{0}^{313}}[1-[1-\frac{\sigma^{2}}{\sigma_{0}^{2}}]^{1/2}\frac{T_{0}^{2}}{T^{2}}]^{313}]]^{2}
\label{a26}
\end{eqnarray}

\begin{eqnarray}
&&  \rho_{I-B}=\frac{3}{8\pi L_{P}^{2}} H_{I-B}^{2} = \nonumber\\&&\frac{3}{8\pi L_{P}^{2}} [\frac{313\pi^{14}k^{14}c^{14}L_{P}^{2}}{T_{D3}^{8}\sigma_{0}^{73}} \times \nonumber\\&&e^{-[1-[1-\frac{\tau^{2}}{\tau_{0}^{2}}]^{1/2}\frac{T_{0}^{2}}{T^{2}}]^{8}[1-[1-\frac{\sigma^{2}}{\sigma_{0}^{2}}]^{1/2}\frac{T_{0}^{2}}{T^{2}}]^{5}}\times\nonumber\\&& \frac{1}{[1-[1-\frac{\tau^{2}}{\tau_{0}^{2}}]^{1/2}\frac{T_{0}^{2}}{T^{2}}]^{73}}\times \nonumber\\&&\frac{1}{[1+ \frac{8\pi^{5}k^{5}c^{5}L_{P}^{2}}{T_{D3}^{313}\sigma_{0}^{313}}[1-[1-\frac{\sigma^{2}}{\sigma_{0}^{2}}]^{1/2}\frac{T_{0}^{2}}{T^{2}}]^{313}]}]^{2}
\label{a27}
\end{eqnarray}

The above results show that by increasing the temperature of the BIon and the acceleration of the system,  the energy density of the universe which lives in part A of the  BIon in region I decreases, while the energy density of the universe in part B of the BIon in region I increases. This means that by increasing the temperature of the system, the energy of one part of an accelerating BIon is going out and entering into another part..

\section{Summary and Discussion} \label{sum}
 We constructed the Padmanabhan idea in an accelerating BIon,  and argued that the birth and expansion of the universe are controlled by the evolution of the BIon in extra dimensions. We have shown that the acceleration of the BIon leads to the difference
between the number of degrees of freedom on the boundary surface of the universe and the number of degrees
of freedom in a bulk region. Also, we have shown that by increasing the acceleration of the BIon, the scale factor and energy density of the universe grow. 

\section*{Acknowledgements}
\noindent The work of Alireza Sepehri has been supported
financially by the Research Institute for Astronomy and Astrophysics
of Maragha (RIAAM), Iran under the Research Project No. 1/5237-79. The authors would like to sincerely thank an anonymous referee for constructive criticism which led to an improvement in the manuscript.

\end{document}